\documentclass[useAMS,usenatbib]{mn2e}

\usepackage{epsfig}
\usepackage{graphicx}
\usepackage{amsmath}
\usepackage{amssymb}
\usepackage{myaasmacros}

\newcommand{\beq}{\begin{equation}}
\newcommand{\eeq}{\end{equation}}

\def\eps{\epsilon}
\def\epsw{\eps_{\rm f}}
\def\Edotw{\dot{E}_{\rm w}}
\def\Lx{L_{\rm X}}
\def\Ewind{E_{\rm wind}}
\def\ergs{\rm erg~s^{-1}}
\def\Msun{M_{\odot}}
\def\Msunh{M_{\odot}h^{-1}}
\def\Mbh{M_{\rm BH}}

\def\Mdotacc{\dot{M}_{\rm acc}}
\def\Mdotinf{\dot{M}_{\rm inf}}
\def\Mdotoutf{\dot{M}_{\rm outf}}
\def\Mdotbh{\dot{M}_{\rm BH}}
\def\Mdotedd{\dot{M}_{\rm Edd}}
\def\vw{v_{\rm w}}

\def\kms{\rm km~s^{-1}}
\def\pdot{\dot{p}}

\title[AGN Feedback in Galaxy Merger]
{Consequences of Mechanical and Radiative Feedback from Black Holes in Disc Galaxy Mergers}

\author[Choi et al.]{\parbox[t]{\textwidth}{
 Ena Choi$^{1}$\thanks{E-mail:echoi@astro.princeton.edu}, 
 Thorsten Naab$^{2}$,
 Jeremiah P. Ostriker$^{1,3}$, 
 Peter H. Johansson$^{4}$, 
 Benjamin P. Moster$^{2}$}
 \vspace*{6pt} \\
$^1$ Department of Astrophysical Sciences, Princeton University, Princeton, NJ 08544, USA \\
$^2$ Max-Planck-Institut f\"ur Astrophysik,
  Karl-Schwarzschild-Strasse 1, 85741 Garching, Germany \\  
$^3$ Department of Astronomy, Columbia University, New York, NY 10027, USA \\
$^4$ Department of Physics, University of Helsinki, Gustaf H\"allstr\"omin katu 2a, FI-00014 Helsinki, Finland}

\begin{document}

\date{Accepted ???. Received ??? in original form ???}
\maketitle
\label{firstpage}

\begin{abstract}
We study the effect of AGN 
mechanical and radiation feedback on the formation of bulge 
dominated galaxies via mergers of disc galaxies. The merging galaxies
have mass-ratios of 1:1 to 6:1 and include pre-existing hot gaseous
halos to properly account for the global impact of AGN feedback. Using 
smoothed particle hydrodynamics simulation code (GADGET-3) we compare three
models with different AGN feedback models: (1) no black hole and no AGN feedback; (2)
thermal AGN feedback; and (3) mechanical and radiative AGN
feedback. The last model is motivated by observations of broad
absorption line quasars which show winds with initial velocities of
$\vw \ge$ 10,000 $\kms$ and also heating associated with the central
AGN X-ray radiation. The primary changes in gas properties due to
mechanical AGN feedback are lower thermal X-ray luminosity from the
final galaxy - in better agreement with observations - and galactic 
outflows with higher velocity $\sim 1000$ $\kms$ similar to recent 
direct observations of nearby merger remnants. The kinetic energy 
of the outflowing gas is a factor of $\sim$ 20 higher than in the 
thermal feedback case. All merger remnants with momentum-based AGN
feedback with $\vw \sim 10,000$ $\kms$ and $\epsw=2 \times 10^{-3}$, 
independent of their progenitor mass-ratios, reproduce the 
observed relations between stellar velocity dispersion and black hole
mass ($\Mbh - \sigma$) as well as X-ray luminosity ($L_X - 
\sigma$) with $10^{37.5}\lesssim L_X (0.3-8~{\rm keV})/\ergs  \lesssim 10^{39.2}$
for velocity dispersions in the range of $120~\kms \lesssim \sigma \lesssim
190~\kms$. In addition, the mechanical feedback produces a much 
greater AGN variability. We also show that gas is more rapidly and 
impulsively stripped from the galactic centres driving a moderate
increase in galaxy size and decrease in central density with the 
mechanical AGN feedback model.
\end{abstract}
\begin{keywords}
accretion, accretion discs -- black hole physics -- galaxies: 
active-- galaxies: nuclei -- galaxies: formation -- quasars: general
\end{keywords}

\section{Introduction}\label{intro}
Accretion onto central massive black holes in galactic nuclei emits
energy in the form of electromagnetic radiation, relativistic  jets,
and wider angle non-relativistic outflows 
\citep{1969Natur.223..690L,1984ARA&A..22..471R}. The
coupling of the energy output to the gas in galaxies, i.e.  
active galactic nucleus (AGN) feedback, is believed to play an 
important role in galaxy formation by regulating central star
formation and quenching cooling flows and produce 
approximately the linear relationship between the central massive 
black hole and the stellar component of the elliptical galaxy 
\citep{1989IAUS..134..217D,1993nag..conf..197K,
1998AJ....115.2285M,2000ApJ...539L..13G,2002ApJ...574..740T,
2003ApJ...589L..21M,2007ApJ...665..120A,2009ApJ...698..198G}. 
However, understanding the precise physical mechanisms by which 
this feedback occurs poses a major challenge for our understanding of 
the connection between AGN physics and galaxy evolution. Several 
mechanisms have been proposed and numerically investigated, 
including radiative heating 
\citep{2005MNRAS.358..168S,2007ApJ...665.1038C,2009ApJ...699...89C}, 
radiation pressure \citep{2010MNRAS.406L..55D},
cavities generated by the injection of thermal energy 
or cosmic rays \citep{2004MNRAS.355..995D,
2007MNRAS.380..877S,2008MNRAS.387.1403S,2010ApJ...712.1311G,
2012ApJ...752...22B,2012ApJ...752...24P,2013MNRAS.428.2966P}, 
thermal energy input \citep{2005Natur.433..604D,2005MNRAS.364.1105S} 
or bipolar mechanical outflows and 
jets \citep{2004MNRAS.348.1105O,2010MNRAS.402..789N,2011ApJ...738...54K,2012MNRAS.420.2221D,
2012ApJ...754..125C,2012MNRAS.424..190G,2013AN....334..394G,2014MNRAS.437.1456B}.

Based on one- and two- dimensional computations,
\citet{2010ApJ...722..642O} 
quantitatively studied the relative importance of the different 
processes in protecting the central black hole from excessive mass 
growth and found mechanical feedback with proper mass and 
momentum injection to be the dominant mode of feedback. Energy 
deposition from a central AGN feedback takes place when the 
accretion rates onto the central black hole are high, e.g. when the 
density of the surrounding gas is high. If the cooling time of the gas
is resolved and is sufficiently short, the gas tends to instantly
radiate away any thermal energy input. Therefore, thermal energy input
is a rather inefficient regulator for black hole growth. Recently 
\citet{2012MNRAS.426..140D} pointed out that the thermal AGN 
feedback implemented with the multiphase star-formation model 
\citep{2003MNRAS.339..289S} has negligible effect since 
the energy deposited to star-forming gas particles is radiated 
away quickly. \cite{2014MNRAS.437.1456B} further confirmed
this showing more limited effect of thermal feedback with the 
expansion of the artificial hole around black hole limited. On the 
other hand, momentum input, which cannot 
be radiated away easily, was found to be
very efficient in  limiting the infall and accretion onto the central
black hole in one- and two- dimensional simulations \citep{2010ApJ...722..642O}. 
It also tends to impart considerably more kinetic energy to the
outflowing gas at a  given accretion rate and efficiency of energy
release.  

In this context, \citet{2012ApJ...754..125C} hereafter CONJ12, introduced and tested
the modeling of mechanical feedback from AGN communicating to the
ambient gas via a bipolar wind in three-dimensional smoothed
particle hydrodynamical (SPH) simulations to verify whether this
form of feedback is able to regulate black hole growth. This treatment
also includes a modified algorithm for the black hole accretion rate
with a Bondi radius criterion and the effect of radiation from the
accreting black hole. We used simulations of isolated discs and of one
equal-mass disc merger to demonstrate that massive, non-relativistic
outflows can indeed regulate the black hole growth. Also, the new
treatment of the AGN feedback results in stronger outflows at higher
velocity (up to $\vw \sim 2,000$ $\kms$), a greater fluctuation level
in both the radiant and wind outflow rates, and lower X-ray
luminosities of hot gas compared to the thermal feedback
treatment that is commonly adopted in many three-dimensional
hydrodynamic calculations 
\citep[e.g.,][]{2005ApJ...620L..79S,2005MNRAS.361..776S,
2005Natur.433..604D,2005ApJ...630..705H,2007MNRAS.380..877S,
2009MNRAS.400..100S,2009MNRAS.398...53B,2009ApJ...690..802J,
2011MNRAS.414..195T,2012MNRAS.420.2662D,2013arXiv1304.0443N}.

In this paper we extend our previous work and study the effect of
radiation and strong winds from AGN on the gas as well as the stellar
component of the host galaxies during mergers of equal- and
unequal-mass disc galaxies. These merging disc galaxies now 
include an extended hot gaseous halo \citep{2011MNRAS.415.3750M} as predicted by 
cosmological hydrodynamical simulations of galaxy formation 
\citep{2002MNRAS.335..799T,2010MNRAS.407.1403C,2013MNRAS.432.3005C},
and as inferred from X-ray observations of normal massive disc 
galaxies \citep[e.g.][]{2011ApJ...737...22A,2013ApJ...762..106A,2013MNRAS.428.2085L}.
Such a component was neglected in
previous AGN feedback studies using idealized simulations. It is,
however, most relevant for a proper treatment of the hydrodynamic
interaction of the AGN wind with the ambient medium and for a more
accurate determination of the X-ray properties. 

We further study the
effect of mechanical AGN feedback on the evolution of the stellar
component of the host galaxy. During active phases of the black hole
large amounts of gas can be removed from the central regions of the
galaxies on short timescales. It has been argued that this process can
trigger a significant dynamical expansion of the stellar component
\citep{2008ApJ...689L.101F,2010ApJ...718.1460F,
2010MNRAS.401.1099H,2012MNRAS.422.3081M,2013MNRAS.tmp.1662D}. 
 This process might
be relevant in the context of recent observational studies indicating
that many massive, passively evolving galaxies at high redshift ($z >
1$) are more compact than local galaxies with the same stellar mass
\citep{2004ApJ...600L.107F,2004ApJ...604..521T,2007MNRAS.374..614L,
2007ApJ...671..285T,2007MNRAS.382..109T,2008A&A...482...21C,
2008ApJ...677L...5V,2009ApJ...695..101D}. While much of
the mass growth and corresponding size growth in the outer parts of
giant ellipticals is potentially driven by accretion of stars in minor
mergers \citep[][but see \citealp{2012ApJ...746..162N,
2012MNRAS.422.1714N}]{2012ApJ...744...63O,2012ApJ...754..115J,
2012MNRAS.425.3119H,2013MNRAS.428..641O,
2013MNRAS.429.2924H,2013MNRAS.431..767B}.
Observations by \citet{2008ApJ...677L...5V,2010ApJ...714L.244S,
2012MNRAS.422.3107S,2013ApJ...771...85V} and others 
also indicate a decrease in 
central densities which does not easily occur in the minor merger 
picture \citep[cf.][]{2013MNRAS.429.2924H}. These observations suggest that in
addition to merging an additional mechanism is needed. We test the
contribution of the AGN driven wind to the galaxy size growth and
the central density decrease. 

The paper is organized as follows. In Section~\ref{model}, we 
provide a brief summary of the simulation code and the initial 
conditions. We also summarize the algorithmic implementation of the
black hole accretion and feedback model. We present our result for the   
equal- and unequal-mass merger simulations in Section~\ref{result}
with a detailed analysis of the effect of the assumed feedback model
(`mechanical' vs. `thermal') and the progenitor mass-ratio on star
formation, black hole growth growth and the properties of the AGN
driven wind. In Section~\ref{result:galaxy} we discuss the impact of
mechanical and thermal feedback on the properties of the merger 
remnant and highlight the effect on the X-ray luminosities as well
as the size and central density of the stellar component. We also
discuss the black hole and X-ray scaling relations. Finally, in
Section~\ref{summary} we summarize and discuss our main results. 
 
\section{Methodology}\label{model}
\subsection{Numerical Code}
We perform the simulations using the parallel 
TreeSPH-code GADGET-3 \citep{2005MNRAS.364.1105S}. The code employs the 
Lagrangian SPH \citep[see][]{1992ARA&A..30..543M}  technique for gas particles 
and solves the equations of motion for the collisionless dark matter 
and star particles.  The sub-resolution modeling for star formation
assumes a two-phase medium of hot and cold gas 
\citep{1977ApJ...218..148M,2003MNRAS.339..289S}, and the stars form from a cold component
embedded in sufficiently dense gas, i.e., 
$n > n_{\rm th}= 0.128$~cm$^{-3}$ with the short-lived stars 
supplying a thermal energy of $10^{51}$~erg to the surrounding gas per 
supernovae (SNe). SN-driven galactic winds are not included in this 
study. We include the radiative cooling for a primordial composition 
of hydrogen and helium \citep{1996ApJS..105...19K} and a spatially uniform 
time-independent UV background radiation field with a modified 
\citet{1996ApJ...461...20H} spectrum. The dimensionless Hubble parameter is 
$h=0.71$ such that the present-day Hubble parameter is 
$H_0 = 71$ km $\rm s^{-1}$ $\rm Mpc^{-1}$.

\subsection{Initial conditions and galaxy parameters}
We simulate binary mergers of equal and unequal mass and the 
initial galaxy disc galaxy models are constructed following
\citet{2005MNRAS.361..776S}. We additionally include a diffuse, rotating hot gaseous   
halo as described in \citet{2011MNRAS.415.3750M,2012MNRAS.423.2045M}. The progenitor galaxies are
composed of a rotationally supported disc of gas and stars, a stellar
bulge and a central black hole embedded in a halo consisting of hot
gas and dark matter. Each galaxy has a virial velocity and radius
$v_{\rm vir} = 160~\kms$, and $r_{\rm vir} = 160 $ $h^{-1}$ kpc
corresponding to a  virial mass of $M_{\rm vir} = 9.53 \times 10^{11}$  
$h^{-1} \Msun$. The \citet{1990ApJ...356..359H} profile dark matter halos are 
constructed with a concentration parameter $c=9$ of the corresponding
Navarro--Frenk--White (NRW) halo \citep{1995MNRAS.275..720N}. The dark matter halo
is then populated with exponential discs with a baryonic mass fraction
of $m_d = 0.041$, with a gas fraction of $f_{\rm gas}=0.2$ and with
the rest being stars. We set the disc scale length $r_d$ using the
\citet{1998MNRAS.295..319M} formalism, assuming that the fractional disc angular
momentum equals the disc mass fraction  $m_d$ for a constant halo spin
of $\lambda = 0.033$ for all models. The vertical scale height $z_0$
of the stellar disc is radially constant and set to $0.2 r_d$. The
black hole at the centre of each galaxy is modeled as a collisionless
sink particle which can accrete gas and the initial seed black hole
masses is $10^{6} \Msun$. 

We model the hot gaseous component as a slowly rotating halo with 
a spherical density profile. The density distribution follows the 
observationallyÊmotivated beta-profile \citep{1976A&A....49..137C,1984ApJ...276...38J,1998ApJ...503..569E}.
It has three free parameters: the central density $\rho_0$, theÊ
core radius $r_c$ and the outer slope parameter beta. 
We adopt $\beta =2/3$ \citep{1984ApJ...276...38J}, $r_c=0.22 r_s$ \citep{1998ApJ...497..555M}
and fix $\rho_0$ such that the hot gas mass within the virialÊradius is 
$M_{gas,halo}$. The hot gaseous halo is rotating around the spin axis of the disc.
The angular momentum of the hot gaseous halo is set by
requiring that the specific angular momentum of the the gas. $j_{\rm
  gas,halo} = J_{\rm gas,halo} / M_{\rm gas,halo} $ is a multiple of the specific
angular momentum of the dark matter halo $j_{\rm DM} = J_{\rm DM} /
M_{\rm DM} $ such that $ j_{\rm gas,halo} = a j_{\rm DM}$. A value of
$a=1$ matches the commonly adopted assumption that there is no
angular momentum transport between the dark matter halo and
the gaseous halo. The angular momentum distribution is then assumed to
scale with the product of the cylindrical distance from the spin axis
$R$ and the circular velocity at this distance: $j(R) \propto R \;
v_{\rm circ}(R)$. The vertical velocity of the gas halo particles is
set equal to be zero.

We set the orbital geometry to be G13 \citep{2003ApJ...597..893N} for our merger 
simulation following \citet{2009ApJ...690..802J}. This geometry corresponds to 
the inclinations $i_{p}=-109, i_{s}=180$ and the arguments of
pericenter $\omega_{p}=60, \omega_{s}=0$ for the primary and  
secondary galaxies, respectively. The galaxies approach each 
other on parabolic orbits. The initial separation of the progenitors
of the 1:1 mergers is  $R_{\rm init}=r_{\rm vir}$ with a pericentric
distance of $r_{\rm peri}=2 r_{\rm d}$, where $r_{\rm vir}= 160 h^{-1}$~kpc is 
the virial radius and $r_{\rm d}= 2.5$ $h^{-1} \rm {kpc}$ is the disc 
scale length. For the unequal-mass mergers the initial separation is 
the mean of the virial radii of the two galaxies. The
pericentric distance is $r_{\rm peri}=2 r_{\rm d,mean}$, the mean disc
scale radius of the two progenitors (see Table~1 for details). 
Every simulation was evolved for a
total of $t=3$~Gyr with the merger taking place at around $t\sim
1.5-2.0$~Gyr. 

\begin{table}
   \begin{center}
   \caption{Orbital Parameters of Initial Condition }
    {
   \begin{tabular}{c|c|c}\hline\hline
Mass Ratio & $R_{\rm init}$$^{a}$ & $r_{\rm peri}$ $^{b}$ \\
  \hline
1:1  & 160.0 & 5.0 \cr
2:1 & 143.5 & 4.45 \cr
3:1 & 135.0 & 4.2 \cr
6:1 & 124.0 & 3.8 \cr
  \hline\hline
   \end{tabular}}
   \end{center}
   \label{tab:orbit}
   \begin{flushleft}
    $^{a}$ The initial separation of the progenitors  \\
    $^{b}$ The pericentric distance of the progenitors\\
   \end{flushleft}  
 \end{table}
 
For all simulations, the primary galaxy is realized with $1.7 \times
10^6$ particles: the halo has $8.0 \times 10^5$ dark  
matter particles and $1.3 \times 10^5$ gas particles, the disc has 
$4.8 \times 10^5$ stellar particles and $1.2 \times 10^5$ gas 
particles, and the bulge has $2.0 \times 10^5$ stellar particles. For 
the secondary galaxies in minor mergers, we scale down the galaxy
masses and particle resolution accordingly in order to maintain equal 
mass resolution. The smaller galaxy in a 2:1 merger, for example, has half
the number of particles than the more massive galaxy (See Table~2). All gas and star
particles have the same mass of $m_{\ast, \rm gas}=6.5 \times 10^4$
$\Msunh$ (we spawn one star particle per gas particle), whereas the
dark matter particles have a mass of $m_{\rm DM}=1.1 \times 10^6$
$\Msunh$. The  gravitational force softening length is $\eps = 66$ pc
for the dark matter particles and $\eps=16$~pc for the gas and star
particles respectively. We also perform a resolution study, with twice 
the mass resolution for the fiducial model with the softening length 
$\eps = 52$ pc for the dark matter particles and $\eps=13$~pc for
the baryonic particle scaled with the square root of the mass ratio 
following  \citet{2001MNRAS.324..273D}. The simulation parameters
are summarized in Table~2.

\begin{table*}
   \begin{center}
   \caption{Galaxy Initial Conditions }
    {
   \begin{tabular}{c|c|c|c|c|c|c|c|c}\hline\hline
Model &$M_{\rm DM}$$^{a}$ &$M_{\rm gas,halo}$$^{b}$ & $M_{\rm gas,disc}$$^{c}$ & $M_{\rm \ast,disc}$$^{d}$ & $M_{\rm \ast,bulge}$$^{e}$& $m_{\rm DM}$$^{f}$& $m_{\ast, \rm gas}$$^{g}$  \\
&$\Msunh$&$\Msunh$&$\Msunh$&$\Msunh$&$\Msunh$&$\Msunh$&$\Msunh$ \\
  \hline
Half-mass Progenitor & $4.4 \times 10^{11}$ &$4.2 \times10^{9}$&$3.9 \times 10^{9}$& $1.6 \times 10^{10}$&$6.5 \times 10^{9}$&$1.1 \times 10^6$ & $6.5 \times 10^4$  \cr
Fiducial Galaxy & $8.8 \times 10^{11}$&$8.5\times10^{9}$&$7.8 \times 10^{9}$&$3.1 \times 10^{10}$&$1.3 \times 10^{10}$ & $1.1 \times 10^6$ & $6.5 \times 10^4$   \cr
High Resolution & $8.8 \times 10^{11}$&$8.5\times10^{9}$&$7.8 \times 10^{9}$&$3.1 \times 10^{10}$& $1.3 \times 10^{10}$& $5.5 \times 10^5$ & $3.3 \times 10^4$   \cr
  \hline\hline
   \end{tabular}}
   \end{center}
   \label{tab:ic}
   \begin{flushleft}
    $^{a}$ dark matter mass,  
    $^{b}$ gas mass in halo,
    $^{c}$ gas mass in disc,
    $^{d}$ stellar disc mass,
    $^{e}$ stellar bulge mass,
    $^{f}$ dark matter particle mass, and
    $^{g}$ stellar and gas particle mass.
   \end{flushleft}  
 \end{table*}

\subsection{The traditional thermal black hole feedback model}
In the widely adopted thermal black hole feedback models 
\citep[e.g.][]{2005MNRAS.361..776S}, the sub-grid accretion rate on scales smaller 
than the resolution is estimated with a Bondi-Hoyle-Lyttleton 
parameterization \citep{1939PCPS...34..405H,1944MNRAS.104..273B,1952MNRAS.112..195B}.  
For gas with density $\rho$, sound speed $c_{\rm s}$ and
velocity relative to the black hole $v$, the mass accretion rate 
onto the central region is given as:
\begin{equation}
\dot{M}_{\rm{B}}=\frac{4 \pi \alpha G^{2} M_{\rm BH}^{2} \rho}
                            {(c_{\rm s}^2+v^{2})^{3/2}},
\label{Bondi}
\end{equation}
where $\alpha$ is a dimensionless parameter, which should be 
set to unity as long as we resolve the physics and scales related
to the Bondi accretion. In the framework of multiphase interstellar
medium, however, the accretion rate may be higher than the 
Bondi accretion rate calculated for star forming gas due to the unresolved
cold phase as noted in \citep{2011MNRAS.413.1158B,2013MNRAS.432.3401G}. 
We therefore use $\alpha = 32$ in this work, as adopted in CONJ12.

It is assumed that the accretion is limited to the 
Eddington rate given by
\begin{equation}
\dot{M}_{\rm{edd}} \equiv \frac{4 \pi G M_{\rm BH} m_{\rm{p}}}
                                                  {\epsilon_{\rm{r}}\sigma_{\rm{T}} c}.
\label{Eddington}
\end{equation}
Here $m_{\rm{p}}$ is the proton mass,  $\sigma_{\rm{T}}$ is the
Thomson cross-section and $\epsilon_{\rm{r}}$ is the radiative
efficiency assumed to be a fixed value of 0.1, adopted from the mean  
value for radiatively efficient \cite{1973A&A....24..337S} accretion onto a 
Schwarzschild black hole. The accretion rate in the standard models 
is then
$\dot{M}_{\rm{acc}}=\rm{min}(\dot{M}_{\rm{B}},\dot{M}_{\rm{edd}})$
with no additional requirement that accreted particles be
gravitationally bound to the central black hole. 

In the thermal feedback model \citep[e.g.][]{2005MNRAS.361..776S,
2005Natur.433..604D,2007MNRAS.380..877S,2009ApJ...707L.184J}, the 
feedback energy from the black hole ${E}_{\rm feed}$ has typically 
been assumed to be some fraction $\epsilon_{\rm{f}}$ of the rest 
mass energy of the accreted matter and couples thermally and 
isotropically to the surrounding gas as,
\begin{equation}
\dot{E}_{\rm feed}=\epsw \dot{M}_{\rm{inf}}c^{2}.
\end{equation}
A fixed value of $\epsw=0.005$ is adopted in many previous studies 
\citep{2005MNRAS.361..776S,2007MNRAS.380..877S}, so that 0.5 percent of the total accreted rest 
mass energy is available as thermal energy which is distributed to the
neighboring $\sim 64$ gas particles weighted by the SPH kernel. In
this approach, neither mass nor momentum is added to the ambient
fluid by the black hole and all accretion energy is added via thermal   
energy.

\subsection{The new mechanical black hole feedback model}\label{sec:modified}

For the simulations with mechanical feedback from the AGN we use 
the model presented in CONJ12 which is briefly reviewed in this 
section. We first calculate the rate of the mass infall onto the black
hole with an ``alternative averaging (AA)'' method using:
\begin{equation}
\dot{M}_{\rm{inf,AA}}= 
\left\langle \frac{4 \pi \alpha G^{2} M_{\rm BH}^{2} \rho }
                            {(c_{\rm s}^2+ v^{2})^{3/2}} \right \rangle,
\label{bondi_AA}
\end{equation}
where angle brackets denote the averaging over the SPH kernel. 
This method for the calculation of the black hole mass does the 
calculation in both time and space on an individual particle basis 
and then averages the results over the neighboring 64 particles in 
order to reduce the dependency on the number of SPH particles.

To avoid the unphysical accretion of unbound gas from outside the
Bondi radius we statistically limit the accretion of mass to the gas
within the Bondi radius. Since the mass distribution of each gas
particle is smoothed with the kernel size, we allow for the full
accretion rate only if the total volume of a gas particle resides  
within the Bondi radius. Otherwise, we reduce the probability of 
being absorbed by the black hole (soft Bondi radius criterion 
(SB), see CONJ12). To account for the time that it takes a particle 
at radius $r_j$ to be accreted, we include the free-fall modification (FF) to the 
accretion probability with an extra factor of
\beq
p_{j,\rm ff} = \frac{ \frac{1}{\tau_j}}
{ \frac{1}{N_{\rm sph}} \sum\limits_{j=1}^{N_{\rm sph}} \frac{1}{\tau_j}},
\label{p_ff}
\eeq
where  $\tau_j = {r_j} / (c_{\rm s,\it j}^2+v_{j}^{2})^{1/2}$ is the free fall time 
and $N_{\rm sph}$ denotes the typical number of smoothing neighboring
gas particles of the black hole. For a full description of the soft
Bondi radius criterion and the free-fall modification, see Figure~1
and section 2.4 of CONJ12.  

Motivated by observations of broad absorption line winds, which 
convey energy, mass and momentum into the surrounding gas with 
velocity $\sim 10,000$ $\kms$ outflows corresponding to a typical 
broad line wind velocity \citep{2003ARA&A..41..117C,
2009ApJ...706..525M,2010ApJ...709..611D}, we included 
these observed AGN winds in our numerical treatment following 
\citet{2010ApJ...722..642O}. In our model, the AGN winds carry a mass given by:
\beq
\Mdotoutf= \Mdotinf - \Mdotacc, \label{eq:Mdot}
\eeq
where $\Mdotoutf$, $\Mdotinf$ and $\Mdotacc$ respectively 
denote the outflowing/inflowing mass rate and the mass rate 
actually accreted onto the black hole. For simplicity we assume
that the wind is launched at a fixed speed $\vw =10,000$~$\kms$.
Then a momentum flux carried by the wind is given as,
\beq
\pdot = \Mdotoutf \vw, \label{eq:pdot}
\eeq
and the kinetic energy rate of the outflow is given as,
\begin{subequations}
    \begin{eqnarray}
\Edotw & \equiv & \epsw \Mdotacc c^2, \label{eq:edotw1} \\
      &=& \frac{1}{2} \Mdotoutf \vw^2,\label{eq:edotw2} 
    \end{eqnarray}
\end{subequations}
where $\epsw$ denotes the feedback efficiency.  We can define the 
dimensionless quantity $\psi$, the ratio of the mass outflow rate to 
the accreted rate as,
\beq
\psi \equiv 2 \epsw c^2 / \vw^2=\Mdotoutf / \Mdotacc,
\label{eq:psi}
\eeq
and we can rewrite the equation for the black hole accretion rate as,
\beq
\Mdotacc = \Mdotinf \frac{1}{1+\psi}\label{eq:Mdot_sol}.
\eeq
As discussed in \citet{2010ApJ...722..642O} and CONJ12, in the presence of 
significant AGN winds, not all of the mass entering the central region 
$\Mdotinf$ actually reaches the black hole. For example, with the 
feedback efficiency typically adopted in the literature, 
$\epsw = 0.005$, and with the fixed wind velocity 
$\vw = 10,000$~$\kms$, only 10 percent of the inflowing mass is 
actually accreted onto the black hole while 90 percent is ejected in a wind.

We calculate the dimensionless quantity $\psi$ for the given feedback
efficiency $\epsw$ and wind velocity $\vw$, and stochastically select
the wind particles from all gas particle attracted into the central
zone by the black hole keeping the fraction of wind particles to the
total inflowing particles as $\psi/(1+\psi)$. To deposit the wind mass
and momentum, we give kicks to the gas particles selected following
the stochastic approach. We set the direction of the wind to be
parallel or anti-parallel to the direction of angular momentum of each
gas particle, if the central black holes are surrounded by a gas disc
this procedure results in a wind perpendicular to the disc plane
\citep{2004ApJ...616..688P}. The emitted wind particles share their momentum with
two other nearby gas particles to reproduce the shock heated
momentum-driven flows. We deposit the residual energy into these three
particles in thermal form so that  the total energy is conserved.
Having momentum share starts the cascade with 
twice the number of particles and it makes it approach the Sedov 
solution faster, and makes us less subject to the problem of having not 
enough resolution to correctly represent a hydrodynamic outflow.

In addition to the mechanical feedback described above, X-ray 
radiation from the accreting black hole can be coupled to the 
surrounding gas according to an approximation described in
\cite{2005MNRAS.358..168S}, as in \cite{2010ApJ...717..708C,2011ApJ...737...26N}; CONJ12. The luminosity flux 
from the two black holes is calculated at the position of each gas 
particle, and the flux is converted to the net volume heating rate 
$\dot{E}$ by adopting the \cite{2005MNRAS.358..168S} formulae that include 
Compton heating and photoionization heating. Note that 
Equation~\ref{eq:Mdot_sol}, not Equation~\ref{bondi_AA}, 
determines the AGN luminosity flux and thus the magnitude of the 
radiation feedback. We also include the electromagnetic
momentum, the radiation pressure from the X-ray flux from the 
black hole by adding a momentum per unit time of 
$\dot{p}=\dot{E}/c$. The added force is directed radially away from 
the black holes.

Finally, instead of limiting the maximum accretion rate to the 
Eddington rate (Equation~\ref{Eddington}), we compute the 
Eddington force acting on the surrounding gas particles, directed 
radially away from the black hole as described in CONJ12 and 
allow this force to act on the gas flow through the hydrodynamic 
equations. Naturally it reduces the inflow and increases the 
outflow but accretion exceeding the Eddington rates can 
occasionally occur.

\section{Comparison of Thermal and Mechanical Feedback models}\label{result}

\begin{table*}\label{table:bh}
  \begin{center}
  \caption{The simulated merger sample: black hole properties}            
  \footnotesize{           
  \begin{tabular}{c c c c c c c c}
  \hline\hline       
Model &  Mass &   $\epsw$${}^{a}$ & log $M_{\rm BH,f}$&  $l_{\rm max}^{\rm BH,eff}$${}^{b}$  & $l_{\rm min}^{\rm BH,eff}$ & $\sigma_{\rm fluc}$${}^{c}$ & log $E_{\rm BH}$${}^{d}$\\
 & Ratio & & $\Msun$ &  &  & &erg\\
 (1) & (2) & (3) & (4) & (5) & (6) & (7) & (8) \\
\hline                
No-AGN & 1:1 & - & - & - &  - & - & -\\
\hline
Th-AGN-50${}^{e}$  & 1:1 & $5 \times 10^{-3}$ &        8.14   &  -0.70   &  -5.98  & 0.27&   60.09\\  
Th-AGN-20 & 1:1 & $1\times 10^{-3}$ &       8.42    & -0.70 &    -4.66  &   0.25 & 59.98\\
Th-AGN-05 & 1:1 & $5\times 10^{-4}$ &       8.73   &   -0.70  &   -5.44  &  0.18 & 59.69\\
\hline
MR-AGN-50 & 1:1 & $5\times 10^{-3}$ &             7.83 &     0.70    & -4.30   &  6.00& 59.79\\
MR-AGN-20${}^{f}$ & 1:1 & $2\times 10^{-3}$    &      8.30    &  1.18   &  -6.11 &4.77  &  59.86\\
M-AGN-20${}^{g}$ & 1:1 &  $2\times 10^{-3}$    &    8.40    &  1.58    & -6.51   & 4.80&  59.96 \\
MR-AGN-05 & 1:1 & $5\times 10^{-4}$ &      9.03  &    1.25 &    -4.37   & 4.90 & 59.98\\
High-MR-AGN-20${}^{h}$ & 1:1 & $2\times 10^{-3}$    &   8.26   &   1.16  &   -4.91 & 4.82 &    59.82 \\
\hline    
Th-AGN-50-2:1 & 2:1 & $5\times 10^{-3}$ &       6.94  &   -0.70   &  -3.98  & 0.20  & 58.89\\     
Th-AGN-50-3:1 & 3:1 & $5\times 10^{-3}$ &      6.80  &   -0.70   &  -4.54   &  0.19 & 58.75\\   
Th-AGN-50-6:1 & 6:1 & $5\times 10^{-3}$ &       6.63   &  -0.70    & -4.04  &  0.21 &  58.58\\  
\hline    
MR-AGN-20-2:1 & 2:1 & $2\times 10^{-3}$ &       7.65  &   -0.21   &  -7.91  & 5.88  & 59.20\\     
MR-AGN-20-3:1 & 3:1 & $2\times 10^{-3}$ &      7.46  &   -0.28   &  -9.31   & 5.30  & 59.01\\   
MR-AGN-20-6:1 & 6:1 & $2\times 10^{-3}$ &       7.22   &  -0.79    & -3.90  & 4.21  &  58.77\\  
\hline
\end{tabular}
}
\end{center}  
\begin{flushleft}
$^a$  AGN feedback efficiency\\
$^b$ $l^{\rm BH,eff} \equiv {\rm log} (L_{\rm BH}^{\rm eff} / L_{\rm Edd})$ 
           where $L_{\rm BH,opt}^{\rm eff}$ is the BH luminosity in the 
           optical band after absorption,  i.e., as it will be seen from 
           infinity. The maximum and minimum Eddington rates are 
           listed in column (5) and (6) respectively.\\
$^c$ The fluctuation level of the mass accretion measured
         following Equation~\ref{eq:fluc} after the final coalescence of 
         black hole.\\
$^d$ Total black hole feedback energy distributed throughout the 
         total simulation time.\\
$^e$ Thermal feedback model with commonly adopted feedback 
         efficiency $\epsw = 5\times 10^{-3}$  \citep[e.g.][]{2005MNRAS.361..776S}. \\
$^f$ Our fiducial model with momentum and radiation 
        AGN feedback with modified black hole mass accretion and 
        feedback efficiency $\epsw = 2\times 10^{-3}$. \\
$^g$ Our fiducial model only with momentum  
        AGN feedback with modified black hole mass accretion.
\end{flushleft}
\end{table*}

\begin{table*}
\begin{center}
\caption{The simulated merger sample: Galaxy properties}             
\footnotesize{             
\begin{tabular}{c c c c c c c c c c c c }
\hline\hline       
  \label{table:galaxy}
Model &  Mass
&   $\epsw$ &  $\sigma_{\rm bul,f}$${}^{a}$ & $R_{\rm eff}$ & log $\Lx$${}^{b}$ &
  log $M_{\rm{wind}}$${}^{c}$ &  log $E_{\rm wind}$${}^{c}$ & log $L_{\rm kin}$${}^{d}$  & log $M_{\ast}$${}^{e}$ & log $E_{\rm SN}$${}^{f}$ \\
 & Ratio&  & $\kms$ & kpc &  $\ergs$ & $\Msun$ &  erg & $ \ergs$ & $ \Msun$ & erg \\
  (1) & (2) & (3) & (4) & (5) & (6) & (7) & (8) & (9) & (10) & (11)\\
\hline               
No-AGN & 1:1 & - &      199.9    &  1.14 &    39.16 &    10.03   &  57.44  &   41.34 &    10.30 &    59.57 \\
\hline
Th-AGN-50  & 1:1 & $5 \times 10^{-3}$ &   187.6    &  1.32    & 39.75     & 9.53 &    57.68  &   41.40  &   10.23  &   59.50 \\  
Th-AGN-20 & 1:1 & $2\times 10^{-3}$ &        180.7   &   1.37    & 38.80 &     9.42  &   57.82   &  41.66  &   10.22  &   59.49\\
Th-AGN-05 & 1:1 & $5\times 10^{-4}$ &          172.7    &  1.51   &  38.39  &    9.40  &   58.01   &  42.00    & 10.18   &  59.45 \\
\hline
MR-AGN-50 & 1:1 & $5\times 10^{-3}$ &           182.4  &    1.30  &   39.30  &     9.31 &    58.67   &  42.12  &   10.18   &  59.44\\
MR-AGN-20 & 1:1 & $2\times 10^{-3}$    &     171.8    &  1.55   &  38.24     &    9.29  &   58.54  &   42.02   & 10.09  &   59.36 \\
M-AGN-20 & 1:1 & $2\times 10^{-3}$    &     160.6     & 1.72   &     38.14   &   9.36   &    58.43  &   42.25  &   10.03 &    59.30  \\
MR-AGN-05 & 1:1 & $5\times 10^{-4}$ &          155.7    &  1.97  &   36.66 &   9.41  &   58.65  &   42.47 &    10.06  &   59.33 \\
High-MR-AGN-20${}^{g}$ & 1:1 & $2\times 10^{-3}$ &184.9    &  1.35  &   38.27   &    9.32    &    58.59   &  41.85   &  10.16   &  59.43 \\
\hline
Th-AGN-50-2:1   & 2:1 & $5\times 10^{-3}$ &  139.9   &   1.48 &    41.07  &     9.34   &  57.65  &   39.17   &   9.97   &  59.24 \\
Th-AGN-50-3:1  & 3:1 & $5\times 10^{-3}$ &   126.4   &  1.54   &  40.84   &     9.34 &    57.73   &  38.59  &    9.85  &   59.12  \\
Th-AGN-50-6:1  & 6:1 & $5\times 10^{-3}$ &   120.3   &  1.52    & 40.50 &   9.05   &  57.40   &  37.09   &   9.73   &  59.00 \\ 
\hline
MR-AGN-20-2:1 & 2:1 & $2\times 10^{-3}$ &       132.5    &  1.80    & 36.63    &  9.46 &    58.36    & 41.82    &  9.79  &   59.06 \\     
MR-AGN-20-3:1 & 3:1 & $2\times 10^{-3}$ &        123.6    &  1.76   &  37.48   &   8.96   &  58.14   &  41.52   &   9.70   &  58.97 \\   
MR-AGN-20-6:1 & 6:1 & $2\times 10^{-3}$ &            119.5   &   1.58   &  37.50    &  8.05  &   57.81  &   41.22  &    9.67  &   58.94\\  
\hline          
\end{tabular}
}
\end{center}  
\begin{flushleft}
$^a$ Initial value of pre-merger primary galaxy 
         $\sigma_{\rm ini} \sim 105$ $\kms$.\\
$^b$ X-ray luminosity measured in 0.3-8 KeV band.\\
$^c$ Total amount of ISM mass lost and wind kinetic energy 
         measured at r =5 kpc from the galactic centre after the final 
         black hole coalescence.\\
$^d$ Mechanical luminosity 
         $L_{\rm kin} \equiv \dot{M}_{\rm wind} v_{\rm wind}^2 / 2$ 
         averaged over 0.2 Gyr by the end of the model evolution. \\
$^e$ Total amount of star formed during the model evolution. \\
$^f$ Total supernova feedback energy distributed during the model 
        evolution.
\end{flushleft}
\end{table*}

We explore the effects of AGN feedback with three types of 
simulations: no black hole and AGN feedback (No-AGN); thermal 
AGN feedback (Th-AGN); and momentum-based mechanical and 
radiation AGN feedback with X-ray heating and radiation pressure 
(MR-AGN). All equal and unequal-mass mergers are run with 
identical merger orbits and initial disc orientations. The simulated
final black hole properties and galaxy remnant properties are given in 
Table \ref{table:bh} and \ref{table:galaxy}, respectively. Note that
we use the model for modified black hole mass accretion
(Section~\ref{sec:modified}) only for the MR-AGN models, the
momentum-based mechanical AGN feedback. For Th-AGN models 
we use the standard mass accretion prescription and 
parameters adopted and studied in the previous studies 
\citep[e.g.][]{2005Natur.433..604D} which produce a broad 
agreement with observational constraints. The number following the model
acronyms indicates the assumed feedback efficiency in units of
$10^{-4}$, i.e. Th-AGN-50 is the thermal feedback model with a
feedback efficiency of $\epsilon_{\mathrm{f}} = 5 \times 10^{-3}$;
MR-AGN-05 is the mechanical feedback model with a feedback efficiency
of $\epsilon_{\mathrm{f}} = 5 \times 10^{-4}$ (see Table
\ref{table:bh}).  The model Th-AGN-50 is identical to the black hole
accretion and AGN feedback model described in \citet{2005MNRAS.361..776S}.

\subsection{Black hole growth and star formation rate}
\begin{figure}
\epsfig{file=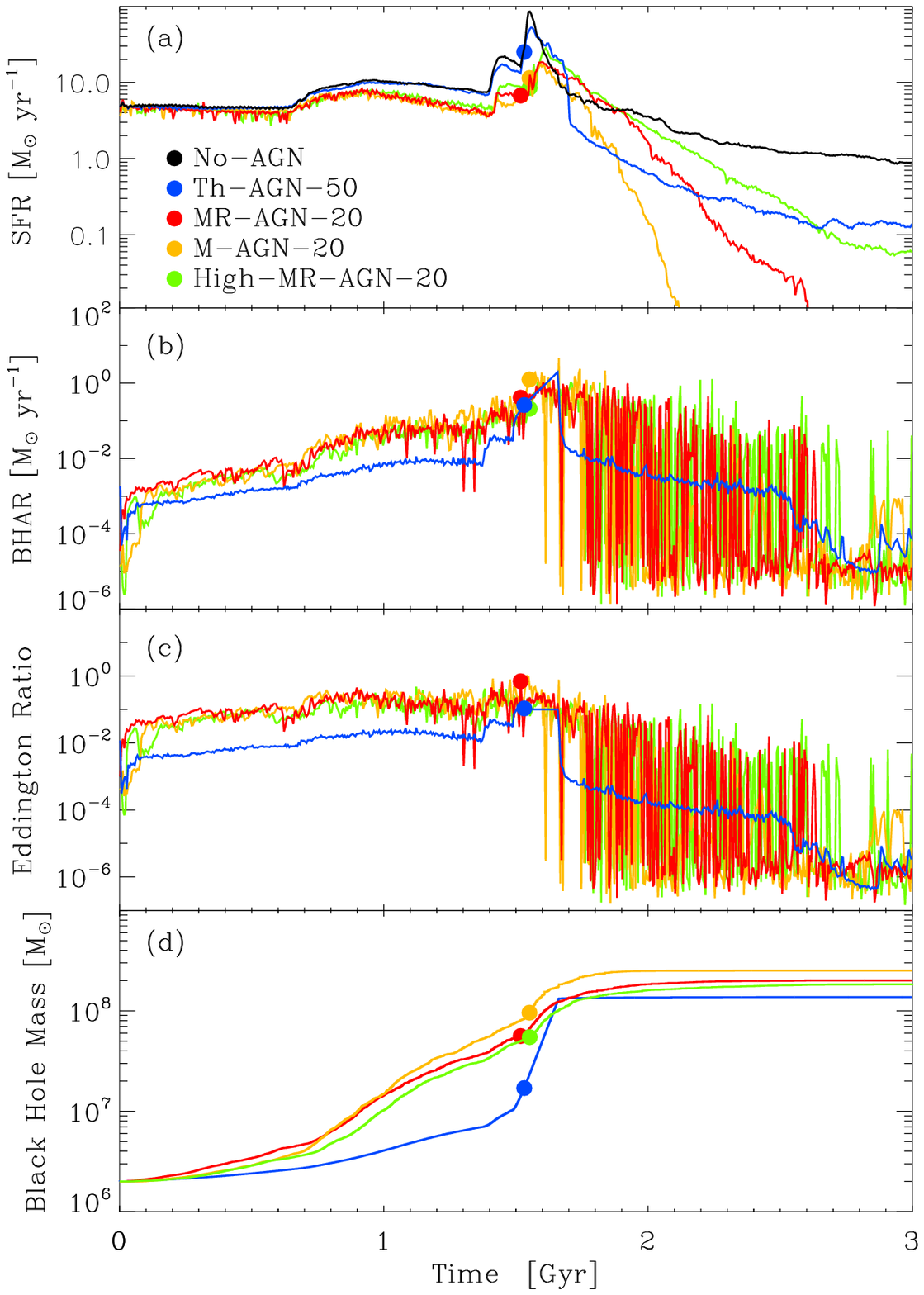,width=\columnwidth}
\caption{
Comparison of the feedback models with a major merger of two 
galaxies:  no AGN (No-AGN, black), thermal feedback (Th-AGN-50, blue), momentum 
and radiation feedback (MR-AGN-20, red), and momentum feedback (M-AGN-20, orange). Higher resolution
run for our fiducial model, High-MR-AGN-20 is shown in green. (a) Evolution of the total star formation 
rate, (b) the total accretion rate onto the black hole, (c) the 
Eddington ratio of the mass accretion ($\Mdotbh/\Mdotedd$), and (d)
the evolution of the black hole mass 
are shown as a function of  
time. The filled circles indicate the time of black 
hole merger. Note that the model outputs are smoothed identically 
to have equal time bin $\Delta t= 5$ Myr.
\label{fig:bh}}
\end{figure}

\begin{figure}
\epsfig{file=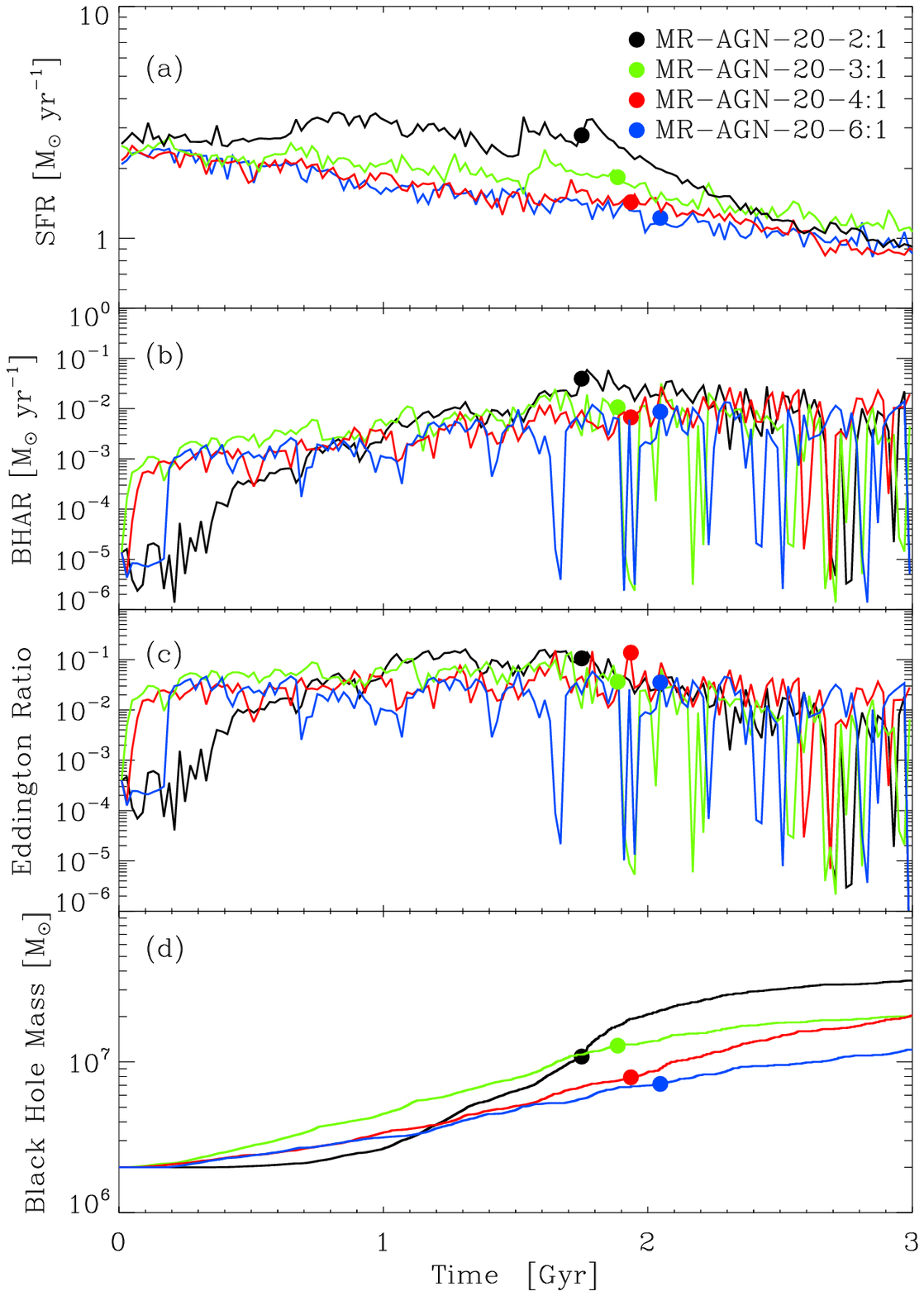,width=\columnwidth}
\caption{(a) The total star formation rate, (b) the total accretion rate 
onto the black hole, (c) the Eddington ratio of the mass accretion 
($\Mdotbh/\Mdotedd$), and (d) the evolution of the black hole mass
as a function of time for 2:1 (black), 3:1 (green), 4:1 (red), and 6:1 
(blue) mergers with the mechanical and radiation AGN feedback. 
The filled circles indicate the time of black hole merger.  The 
model outputs are smoothed identically to have equal time bin 
$\Delta t= 20$ Myr.
\label{fig:bh_minor}}
\end{figure}

Figure~\ref{fig:bh} shows the global star formation rate, the net 
accretion rate onto the black hole, the Eddington ratio of the mass 
accretion ($\dot{M}_{\rm acc}/\dot{M}_{\rm Edd}$) 
and the evolution of the black hole mass (summed over both
black holes prior to merging) for the three different feedback models:
No-AGN (black curve), Th-AGN-50 \citep[e.g.,][blue curves]{2005MNRAS.361..776S},
and MR-AGN-20 (red curves). Note that the thermal feedback model
  Th-AGN-50  adopts a feedback energy coupling efficiency of  
$\epsw = 5\times 10^{-3}$, the value adopted in \citet{2005MNRAS.361..776S} and 
the MR-AGN-20 adopts $\epsw = 2\times 10^{-3}$. We decided to compare
these two models as they result in merger remnants with comparable
final black hole mass. We additionally show the control run only with
the mechanical feedback, M-AGN-20 (orange curves), to quantitatively
show the effect of radiative feedback. We also show the high resolution
run with twice the mass resolution for the fiducial model in green curves
(High-MR-AGN-20).

Figure~\ref{fig:bh}(a) shows that the inclusion of the AGN feedback in
mechanical form reduces the star formation rate before and at
the coalescence of the two galaxies at $t \sim 1.5$ Gyr. After 
the encounter, both AGN  feedback models have lower star formation
rate compared to the no black hole model \citep{2005MNRAS.361..776S}. 
The star formation is more efficiently terminated by the MR-AGN 
feedback than in Th-AGN feedback as the thermal energy added to 
the star forming gas particle is radiated away quickly because of 
the short cooling time. Total amount
of stars formed and the total supernova feedback energy distributed 
during the model evolution for all models are listed in Column (10) 
and (11) of Table \ref{table:galaxy}. The total amount of stars formed
and the corresponding  supernovae feedback energy are reduced by 40 percent
in MR-AGN-20 compared to No-AGN. High resolution run shows 
higher late-time star formation rate compared to the fiducial run resulting in
20 percent more total amount of stars formed due to
the effective cooling and refueling of the gas from gaseous halo.

Figures~\ref{fig:bh}(b) and (c) show the total black hole accretion
rates and the Eddington ratios, respectively. In the Th-AGN-50 model,
the black hole mass accretion rate increases rapidly during the
encounter and final coalescence. In a short time the black hole grows
by an order of magnitude reaches its final mass to within a factor of
a few (Figure~\ref{fig:bh}(d)). Afterwards, the accretion rates are low
with Eddington ratios slowly decreasing from $\dot{M}_{\rm BH}/\dot{M}_{\rm Edd}
\sim 10^{-4}$ to $10^{-5}$. In MR-AGN-20 model, accretion rates are
much more variable with short episodes of efficient accretion even
after the black hole coalescence reaching $\dot{M}_{\rm BH}/\dot{M}_{\rm Edd} \ge
0.1$ (Figure~\ref{fig:bh}(c)). In order to quantify the fluctuation level
of black hole mass accretion for each model, we define a fluctuation
parameter $\sigma_{\rm fluc}$ as,
\beq\label{eq:fluc}
\sigma_{\rm fluc}^2 \equiv \left\langle \left( l_{\rm 1Myr} - l_{\rm 50 Myr}   \right)^2 \right \rangle,
\eeq
where $l_{\rm 1Myr}$ denotes the logarithmic Eddington ratio of black
hole mass accretion measured in time bins of one Myr, and $l_{\rm
  50Myr}$ is the smoothed Eddington ratio measured in 50~Myr time bins
as  $l_{\rm 50Myr}= {\rm log}(\left\langle \dot{M}_{\rm BH}\right
\rangle_{\Delta t=\rm 50 Myr} / \dot{M}_{\rm Edd})$. We determine the global
fluctuation level in the Eddington ratio by averaging over all one 
Myr bins after the black hole merger. The calculated  
fluctuation parameters $\sigma_{\rm fluc}$ are listed in Column (7) in 
Table \ref{table:bh}. The fluctuation level of mass accretion in
MR-AGN-20 is $\sim 4.7$, which is significantly larger than for all
Th-AGN models ($\sim 0.2$). Despite the dramatic differences in the
time dependence of the accretion rates (cf. Figure~\ref{fig:bh}(b)
and (c)), the final black hole masses in the two cases are quite
similar (Figure~\ref{fig:bh}(d)). Including the radiative feedback shows
moderate impact on the growth of black hole, reducing the final
black hole mass by 20 percent. In the high resolution run
we find a slight trend of lower mass accretion rates during the 
initial passage, but the effects are less than 10 percent.

We now proceed to study the star formation histories and the black 
hole accretion histories for mergers with varying mass ratios. We ran
four mergers with mass ratios of 2:1, 3:1, 4:1 and 6:1 on similar
orbits (G13) as the equal mass-merger, adopting the MR-AGN feedback
model. In Figure~\ref{fig:bh_minor}, we show the evolution of (a) the
resulting star formation rates, (b) the total black hole accretion rates,
(c) the Eddington ratios, and (d) the total black hole mass for the
four minor mergers as a function of time. Note that we smooth  
the model outputs shown in Figure~\ref{fig:bh_minor} with longer 
time step ($\Delta t= 20$ Myr) than in Figure~\ref{fig:bh} ($\Delta t= 5$ Myr) to give the overall mean 
evolution. Almost independent of the progenitor
mass-ratio the combined star formation rates are only mildly
decreasing from 2 $M_\odot$/yr to 1 $M_\odot$/yr over 3 Gyrs with only
a mild peak in the 2:1 case during the coalescence. The dependence of 
the mass-ratio is even weaker here than for thermal feedback
models \citep[See Figure 6 in][]{2009ApJ...690..802J} as we include hot gaseous halos which constantly
supply gas \citep{2011MNRAS.415.3750M,2012MNRAS.423.2045M}. A similar
weak evolution and weak trend with mass-ratio is found for the the black
hole accretion rates and the Eddington ratios (Figure~\ref{fig:bh_minor}
(b) and (c)). In all cases, the black hole accretion rates are variable
with the Eddington ratios ranging from $10^{-6}$ to 0.1 (Note the
maximum ratio for equal-mass mergers is about unity) as shown in
Figure~\ref{fig:bh_minor}(c). The evolution of the black hole
accretion rates are mirrored in the growth of the black hole masses,
with the final black hole masses being slightly lower in remnants of
higher mass-ratio  mergers. 

\subsection{Properties of the post merger gas outflow}
\begin{figure}
\epsfig{file=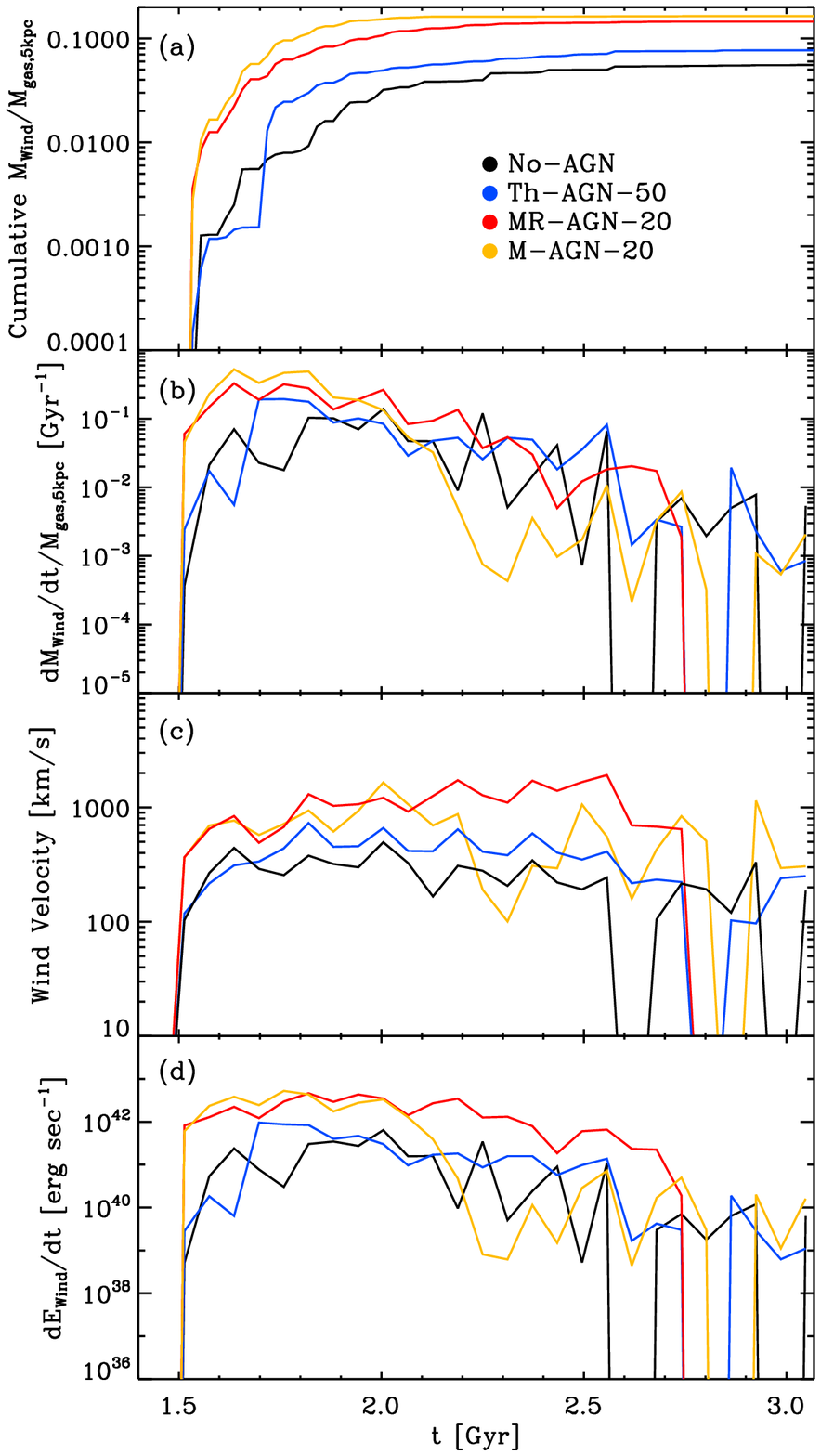,width=\columnwidth}
\caption{The properties of the outflowing gas in three feedback models: 
no AGN (No-AGN, black), thermal feedback (Th-AGN-50, blue), momentum 
and radiation feedback (MR-AGN-20, red), and momentum feedback 
(M-AGN-20, orange) (a) the fraction of mass depleted by
wind within a central sphere with a fiducial galactocentric radius of 
$r=5$ kpc after the final black hole coalescence, (b) the specific 
outflowing mass loss rate, i.e., the rate of gas mass loss normalized by 
the total amount of gas left in the central region of the galaxy
at the final black hole coalescence, (c) outflowing gas velocities and (d) the 
corresponding mechanical luminosities
are shown. 
\label{fig:wind}}
\end{figure}

\begin{figure}
\centering
\epsfig{file=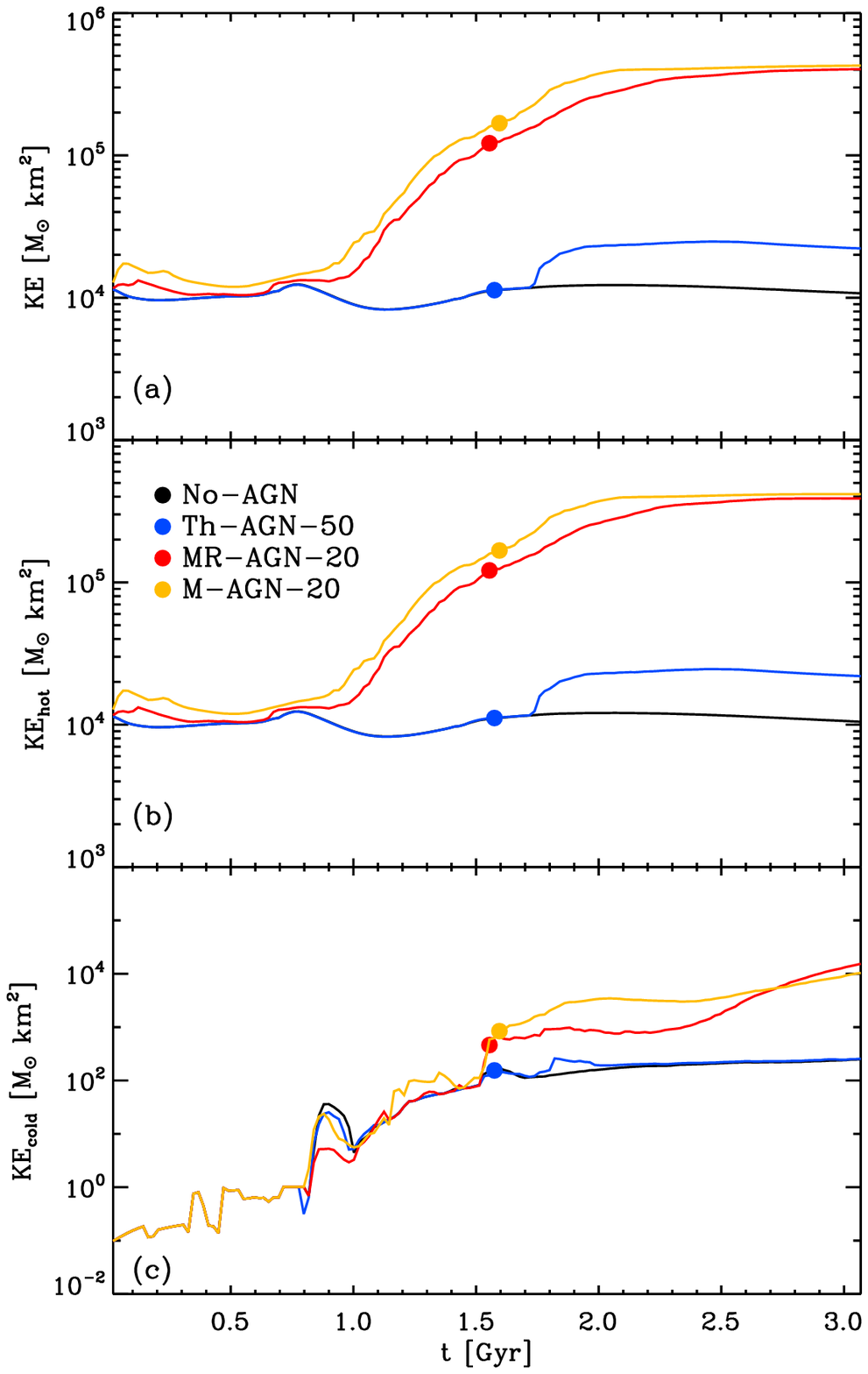,width=\columnwidth}
\caption{(a) Total kinetic energy of the gas, (b) the kinetic energy 
of the warm gas with $T > 10^{4.5}$ K, and (c) the kinetic energy of the
cold gas with $T < 10^{4.5}$ K in the outer region of the
galaxy ($r > 20$ kpc)  are shown for the three 
feedback models: no AGN (No-AGN, black), thermal feedback 
(Th-AGN-50, blue), momentum and radiation feedback (MR-AGN-20, red), 
and momentum feedback (M-AGN-20, orange).
\label{fig:ke}}
\end{figure}

In this section we investigate the physical properties of the gas
outflow from the central parts of the galaxies produced by the three
feedback models. As detailed before, the energy injection conditions
near the black hole in the mechanical and the thermal model are quite
different. In the MR-AGN model, gas surrounding the black hole is
ejected in a wind with an initial velocity of $\sim 10,000$ $\kms$ 
\citep[][CONJ12]{2010ApJ...722..642O}. On the other hand the thermal heating from
the accreting black hole in the Th-AGN model drives slow and hot
outflows in the vicinity of the black hole as shown in \citet{2005MNRAS.361..776S}. 

To parametrize the cumulative mass outflow from the central parts of
the merger remnants we measure the fraction of gas mass leaving a 
central sphere with a fiducial galactocentric 
radius of $r=5$ kpc after the final black hole coalescence. 
We also measure the instantaneous outflow rate
and the velocity of the outflowing gas. With this we calculate a
corresponding mechanical luminosity as $L_{\rm wind} \equiv
\dot{M}_{\rm wind} v_{\rm wind}^2 / 2$, i.e., the kinetic energy
carried away by the outflowing gas.

The temporal evolution of the outflowing gas
properties (the cumulative outflowing gas mass leaving
a central region normalized by the total amount of gas
in the central region at the final black hole coalescence, 
the specific outflow rate, i.e., the rate of gas mass loss
normalized by the total amount of gas left at the final black
hole coalescence, outflow velocity and mechanical luminosity) after the final
black hole coalescence at $t \sim 1.5$
Gyr of the No-AGN model and our fiducial AGN feedback models Th-AGN-50,
MR-AGN-20 and M-AGN-20 are shown in Figure~\ref{fig:wind}. 

The MR-AGN-20 model depletes the gas in the central region
of the galaxy most effectively and loses the largest fraction of gas after the
merger, i.e., $\sim 15$ percent of gas leaves a central sphere by outflow (Figure~\ref{fig:wind}(a)).
The model without black hole also loses the large amount of gas 
after the merger, primarily driven by the merger induced shock and
star formation, but the fraction of outflowing gas mass to the total
gas mass within a central sphere is smaller compared to the AGN
feedback models.

The corresponding specific outflowing mass loss rates are shown
in Figure~\ref{fig:wind}(b). The MR-AGN-20 model has the highest
rate, especially after the merger with the timescale for gas depletion
by outflow of $\tau_{\rm depletion} \sim 2.9$ Gyr. At later times the 
specific outflowing mass loss rate for the mechanical AGN model 
drops and becomes comparable to other models. 

The MR-AGN-20 model
also shows the highest outflow velocities of $\vw \sim 800-1500$ 
$\kms$ (Figure~\ref{fig:wind}(c)). These values are comparable to
recent observations which indicate outflow velocities in the range 
of $700$ km/s to $3000$ km/s \citep{2011ApJ...727...71F,
2011MNRAS.413.1251P,2011ApJ...733L..16S,
2011ApJ...739...69M,2013ApJ...768...75R}. The velocities in the thermal 
AGN model are much lower and do not exceed the $400-500$ $\kms$ which
are comparable to the velocity of the shock heated gas in the No-AGN model. 

The mechanical luminosities of the outflowing gas for the three
models are shown in Figure~\ref{fig:wind}(d). Overall, 
Th-AGN-50 has lower mechanical luminosity $L_{\rm wind} \sim 
10^{41}\rm$ erg/s, mainly because of its slow outflow velocity. The 
MR-AGN-20 with momentum feedback has mechanical luminosities a 
factor of 5-10 higher than the thermal feedback. The total kinetic 
energy carried away by the winds, i.e., the mechanical 
luminosities integrated over the simulation time after the merger for 
the Th-AGN-50 model is $\Delta \Ewind \sim 1.94 \times 10^{58}$
ergs, while the MR-AGN-20 model deposits 
$\Delta \Ewind \sim 6.89 \times 10^{58}$~ergs (3.6 times larger) 
into the ISM within 1.5 Gyrs. The properties of the outflowing gas
of the control run without X-ray radiative heating (M-AGN-20) are
essentially similar to our fiducial model except for the higher
cumulative wind mass. This is due to the higher black hole mass
growth in M-AGN-20.

The outflowing gas removed from the central region would propagate
to the outer region of the galaxy and increase the kinetic energy of
the gas component. Figure~\ref{fig:ke}(a) shows the total kinetic energy of the gas in the
outer region of the galaxy ($r>20$ kpc). In the MR-AGN-20, the bulk 
of mass and kinetic energy are dissipated by the outflowing gas from 
the AGN and located in the outer part of the galaxy. 
The Th-AGN-50 also distributes increased kinetic energy
in the outer part,  especially right after the final coalescence of
the black holes, but the increment is much smaller compared to the
MR-AGN model with the AGN-induced winds. The total kinetic
energy in the outflow, as seen in Figure~\ref{fig:ke}(a),
is a factor of  20 higher in the MR-AGN then in the Th-AGN case. While
the kinetic energy in the outer region of the galaxy is dominated by
the warm component with the temperature $T>10^{4.5}$ K, as shown in
Figure~\ref{fig:ke}(b), the MR-AGN model also  
sweeps up and drives out cold gas increasing the kinetic energy of 
the cold component with $T<10^{4.5}$ K (see Figure~\ref{fig:ke}(c)).
The Th-AGN model, however, has negligible effect on the cold gas 
component within the framework of the standard multiphase star
formation model as pointed out by \cite{2014MNRAS.437.1456B}.
We find a slight trend of higher kinetic energy during the 
initial passage in the mechanical feedback run without X-ray 
heating (M-AGN-20) but this is originated from the different
black hole mass accretion history, higher black hole mass and 
black hole accretion rate in M-AGN-20. 

\section{Properties of merged galaxies}\label{result:galaxy}
\subsection{Distribution of gas during the merger}
\begin{figure*}
\centering
\epsfig{file=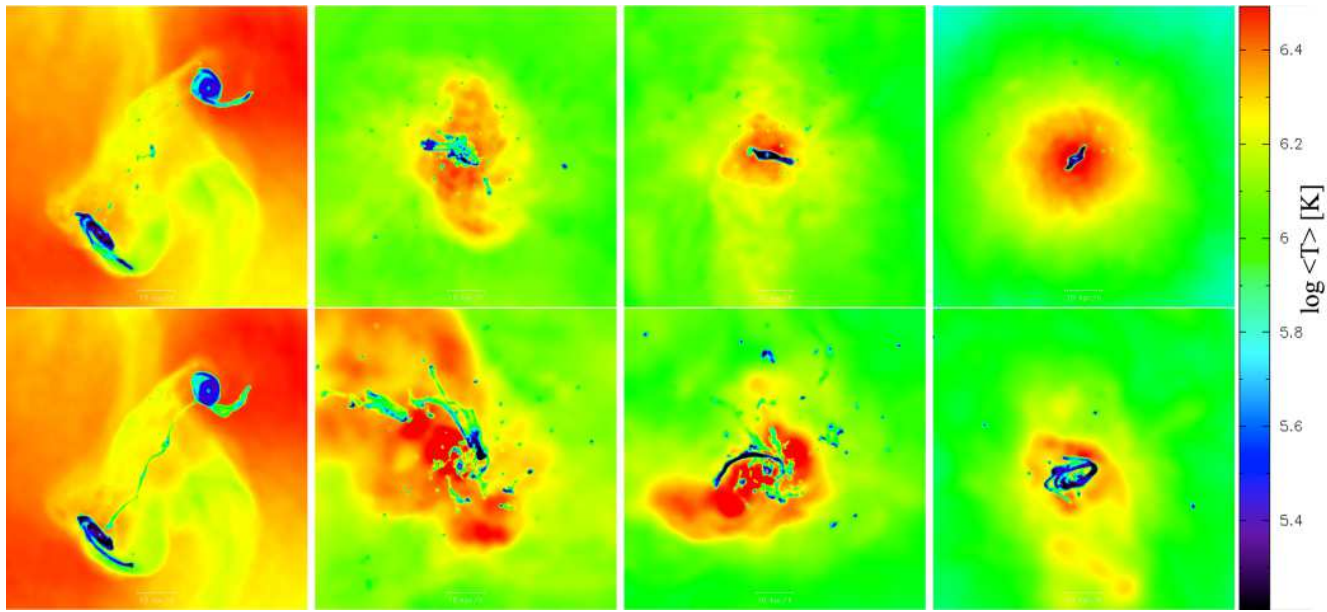,width=\textwidth}
\caption{A sequence of snapshots of the mass-weighted gas 
temperature during a major merger of two models: Th-AGN-50 (top panels) and 
MR-AGN-20 (bottom panels). The snapshots after the first close 
passage of the two galaxies at $t=0.92$ Gyr, right after the galaxies 
and black holes merge at $t=1.78$ Gyr, and afterwards at 
$t=2.0$, 2.64 Gyr are shown from left to right. The images are 
80~kpc on a side and redder color indicates a higher temperature. 
\label{fig:snapshot}}
\end{figure*}

Figure~\ref{fig:snapshot} shows a sequence of snapshots of
the mass-weighted average gas temperature along the line
of sight during a merger of
Th-AGN-50 (top) and MR-AGN-20 (bottom). The 
snapshots show the first close passage at $t=0.92$ Gyr, the time right after 
the final coalescence at $t=1.78$ Gyr and afterwards at $t=2.0$, 
and 2.64 Gyr from left to right. In Th-AGN-50 model,
the heated gas expands  from the central region and forms a hot gaseous halo 
by the end of the simulation while the MR-AGN-20 model show less hot gas. In both simulations we found tidal condensations 
which consist of cold dense gas but there are significantly more materials 
at large distance in dense cold blobs in MR-AGN-20 model. 
\subsection{X-ray properties}
\begin{figure}
\epsfig{file=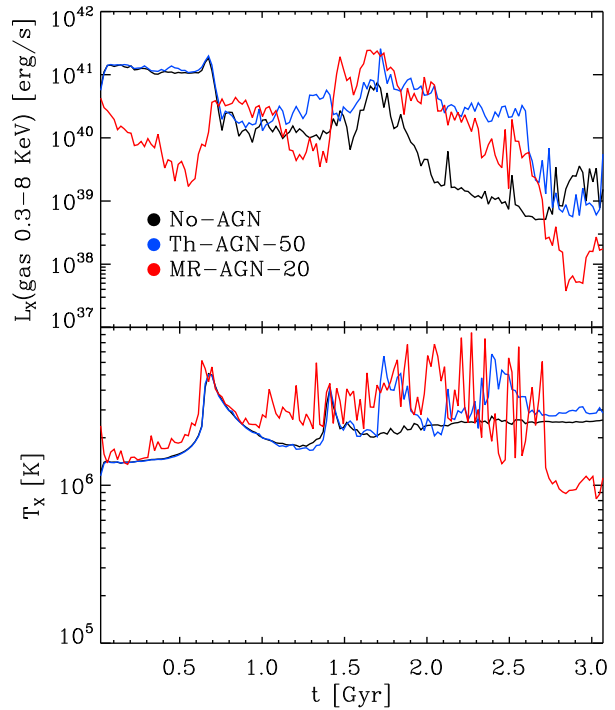,width=\columnwidth}
\caption{Galactic X-ray coronal luminosity $L_X$ (top), and the
X-ray luminosity-weighted temperature of the hot and diffuse gas 
component with a temperature of $T \geq 10^6$ K, and a density 
$\rho \leq 3.16 \times 10^{-3} \Msun$ $\rm pc^{-3}$ (bottom).
\label{fig:lx}}
\end{figure}

We compare the evolution of hot gas and X-ray emission of the 
No-AGN and two feedback models. 
Following \cite{1995ApJ...451..436C},
we calculate the X-ray luminosity  due to bremsstrahlung radiation 
as well as line emissions of all the species using the computed
X-ray emissivity spectra tables, kindly made available to us by 
R. Cen. The emission rates from H, He and metals are given
separately assuming that the gas is in ionization equilibrium
and optically thin. We calculate the total emissivity by summing 
the emission rates of the all components weighted by the 
relative abundance assuming solar abundance. 
Then we compute the integrated luminosity 
for 0.3-8 keV band using the computed total emissivity.
We assume that the central region of the galaxy remains obscured 
because of the large column density of intervening gas and dust, 
therefore we only include the X-ray contribution from hot and diffuse gas 
particles. Following \cite{2006ApJ...643..692C}, we define the 
`hot and diffuse gas' with a temperature 
of $T \geq 10^6$ K, and a density $\rho \leq 3.16 \times 10^{-3} 
\Msun \rm pc^{-3}$, which corresponds to the critical density for star 
formation, so effectively we exclude star forming gas.

Figure~\ref{fig:lx} shows the X-ray luminosities for photons with 
energies of 0.3-8 keV for No-AGN, Th-AGN-50 and MR-AGN-20 as a function of time.
Before the galaxy interaction, the majority of the X-ray emission 
is produced by pre-existing hot halo gas, but in the MR-AGN-20 
model, the high velocity gas outflow by AGN momentum feedback 
effectively drives out the gaseous halo, lowering the X-ray luminosity. 
The system begins to emit more X-rays at the first interaction of 
galaxies at $\sim 0.7$~Gyr, in shocks that lie directly between the 
two discs. During the galaxy interaction, the majority of the X-ray 
emission is produced by shock-heated gas, as Th-AGN-50 shows very 
similar X-ray luminosity compared to No-AGN model. By the time of 
the final coalescence, $t \sim 1.5$ Gyr, however, the X-ray luminosity 
increases in both AGN models.  In the Th-AGN-50 model, the X-ray 
luminosity decreases much more slowly as the black hole keeps 
depositing a significant amount of thermal energy.  On the other hand, in the 
mechanical feedback model, the hot gas in the remnant is driven 
outwards more effectively and the X-ray luminosity decreases 
quickly. The final X-ray emission of the Th-AGN-50 is higher 
($1.6 \times 10^{39}$ $\ergs$) compared with the observed
luminosity (0.3-8 KeV) of the galaxies with corresponding velocity
dispersions. The MR-AGN-20 model, however, results in lower
X-ray luminosity ($9.8 \times 10^{37}$ $\ergs$) that better 
reproduces observed values \citep[e.g.,][]{2011ApJ...729...12B}.

\subsection{Galaxy-black hole Scaling relation}\label{result:scale}

\begin{figure}
\epsfig{file=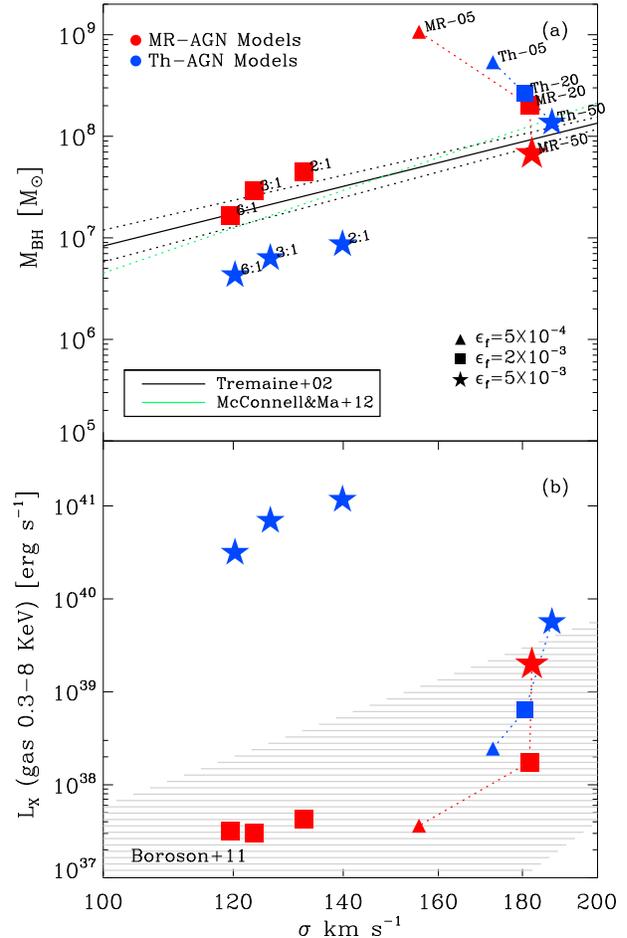,width=\columnwidth}
\caption{ (a) $\Mbh - \sigma$ relation of the two feedback models.
The blue symbols show Th-AGN models and red symbols show the 
MR-AGN models. Different symbols indicate the different feedback
efficiencies. The merger progenitor mass ratios are indicated for
the minor merger cases.
The black and green lines show the observed relation with errors by
Tremaine et al. (2002) and by McConnell \& Ma (2013).
(b) X-ray luminosity of the 
hot gas is plotted against stellar velocity dispersion $\sigma$. 
Observed relation (hatched region) is from Boroson et al. (2011).
\label{fig:relation}}
\end{figure}

We now proceed with a detailed analysis of the effect of the black 
hole feedback prescription and feedback efficiency on the final 
mass of black holes and the corresponding stellar velocity 
dispersions in the merger remnants. We simulate equal-mass 
mergers with different feedback efficiencies and unequal-mass 
mergers with the adopted fiducial feedback efficiencies 
($\epsw=5 \times 10^{-3}$ for Th-AGN and 
 $\epsw=2 \times 10^{-3}$ for MR-AGN) for both feedback models 
 and compare them with the observed $\Mbh - \sigma$ relation.
We calculate the final black hole mass together with the 
mass-weighted line-of-sight stellar velocity dispersion $\sigma$ 
measured from all stellar particles within the projected half-mass 
radius $r_{\rm e}$. The feedback efficiency has a strong effect on 
the final black hole mass of the merger remnant in both feedback 
models, with the higher efficiency producing lower black hole 
masses. For the final stellar velocity dispersion on the other hand, 
the effect of feedback efficiency is less pronounced. The simulated
Th-AGN-50 and MR-AGN-20 results are overplotted on the 
observed $\Mbh - \sigma$ by \cite{2002ApJ...574..740T} and \cite{2013ApJ...764..184M} as 
shown in Figure~\ref{fig:relation}(a). In unequal-mass merger cases, 
MR-AGN-20 models evolve close to the observed $\Mbh - \sigma$
relation while Th-AGN-50 models result in lower black hole mass 
compared to the observed relation.

We show the X-ray luminosity of the hot gas against the stellar
velocity dispersion $\sigma$ in Figure~\ref{fig:relation}(b). 
Observationally all galaxies 
with a shallow potential well with $\sigma < 200~\kms$ seem to 
have only a small amount of hot gas with $L_X < 10^{40}~\ergs$ 
\citep{2011ApJ...729...12B}. Observed $L_X$(gas) is correlated with $\sigma$, although not as 
strongly as the $\Mbh - \sigma$ relation. The effect of the black hole feedback prescription and 
feedback efficiency on the final stellar velocity dispersion is minor 
resulting in typically $160 < \sigma < 190$ $\kms$, however the effect
on the X-ray luminosity is significant. Adopting higher feedback 
efficiency produces higher X-ray luminosity. Th-AGN-50 produces 
X-ray luminosities that are higher for the corresponding velocity 
dispersions. We show that reducing the feedback efficiency to 
$\epsw=5 \times 10^{-4}$ produces X-ray luminosity within the 
observed range, but the resulting final black hole mass is factor of 
20 larger than the one for its corresponding $\sigma$ in the thermal
feedback cases. On the other 
hand, MR-AGN models result in lower X-ray luminosity compared
to the Th-AGN models, and our proposed model MR-AGN-20
reproduces both the observed $L_X - \sigma$ relation and 
$\Mbh - \sigma$ relation. In case of minor mergers, all Th-AGN 
models show excessive X-ray luminosity compared to the 
observed range. The physical reason for the different results is 
easy to understand. The MR-AGN models tends to expand the 
gaseous halo more than the Th-AGN, thus reducing the gas 
density and the thermal X-ray luminosity.

\subsection{Galaxy Size}\label{sec:size}
\begin{figure}
\epsfig{file=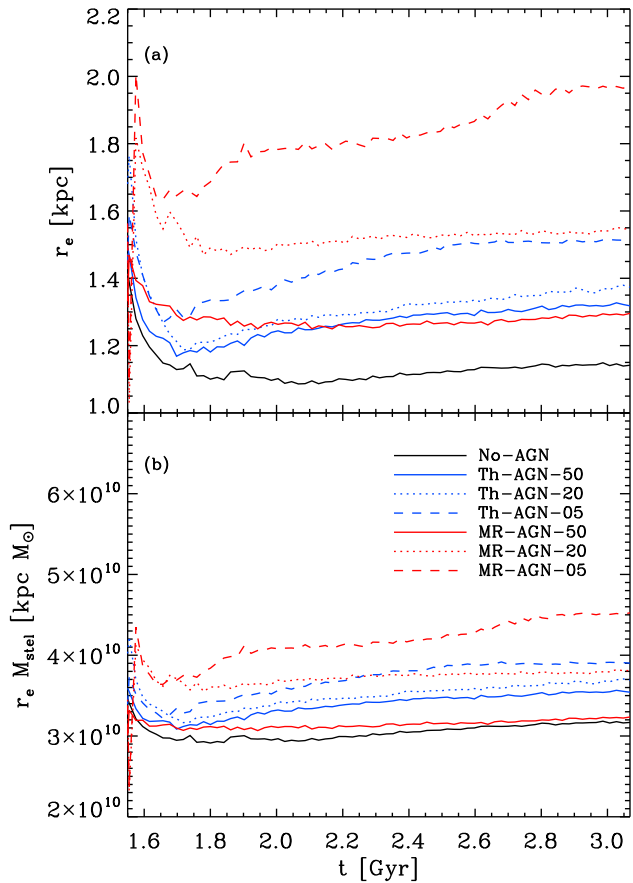,width=\columnwidth}
\caption{(a) Evolution of the effective radius, and (b) the adiabatic 
invariant for two feedback models. 
\label{fig:re}}
\end{figure}

Observations have shown that quiescent massive galaxies at high 
redshift are much more compact than the local galaxies with the 
comparable mass and they have grown in physical size from 
$z\sim1$ to $z \sim0$ \citep[e.g.,][]{2004ApJ...600L.107F,2004ApJ...604..521T,2007MNRAS.374..614L,
2007ApJ...671..285T,2007MNRAS.382..109T,2008A&amp;A...482...21C,
2008ApJ...677L...5V,2009ApJ...695..101D}. Some physical explanations have been 
suggested to explain this growth in size while avoiding the
overproduction of present-day massive galaxies. Minor mergers
involving lower mass galaxies has been proposed as one of the
explanations, as they can produce efficient size growth 
\citep{2009ApJ...699L.178N}. \cite{2010MNRAS.401.1099H} considered the various proposed 
channels for observed size evolution, and concluded that minor dry 
mergers are the ``prime candidate'' for explaining the majority of the
observed sized growth, though other channels including adiabatic 
expansion due to mass loss from stellar winds should play a 
non-negligible role. However, the accretion of satellite stellar systems 
in minor mergers mainly contribute to the mass growth in the outer 
parts of elliptical galaxies \citep{2012ApJ...744...63O,2013MNRAS.429.2924H} and in this picture a 
factor of two decrease in central densities observed by 
\cite{2007ApJ...669..184A,2008ApJ...677L...5V,2010ApJ...714L.244S} 
cannot be fully explained \citep[cf.][]{2013MNRAS.429.2924H}. 
Other candidates include AGN feedback-driven star formation 
\citep{2013MNRAS.431.2350I} and secular processes such as adiabatic expansion 
driven by the expulsion of a substantial fraction of the gas out of the 
galaxy by stellar winds and/or strong AGN feedback \citep{2008ApJ...689L.101F,2010ApJ...718.1460F}.
In the recent studies \citep{2012MNRAS.422.3081M,2013MNRAS.tmp.1662D}, 
it is shown that the galaxies simulated with AGN feedback is more extended 
due to the AGN feedback induced gas expulsion compared to the no-AGN cases.

In this section, we test the possible contribution of the {\it puffing-up}
process by the AGN mechanical feedback prescriptions, which have very
different timescales for the AGN-driven winds. We also check whether
the size increase by the AGN feedback induced mass loss which can 
account for the observed evolution is permitted by observational
 constraints on $\Mbh - \sigma$.

The effect of the mass loss on the structure and dynamics of a 
stellar system depends on the amount of mass loss and on the 
timescale of ejection. The puffing up of a virialized stellar system 
by rapid mass loss is a well-known phenomenon, extensively 
studied both analytically and through numerical simulations, with 
reference to galaxies \citep{1979ApJ...230L..33B}, and to star clusters 
\citep{1980ApJ...235..986H,2011MNRAS.414.3690R}. Adiabatic expansion can be also caused by 
much slower mass loss, for example, by stellar winds. If we define
the fraction of changes in radius and in mass respectively as
$\delta_r \equiv (r_1 - r_0)/r_0$ and 
$\delta_m \equiv (m_1 - m_0)/m_0$ where $m_0$ and 
$m_1$ are the initial and final masses and $r_0$ 
and $r_1$ are the initial and final radii, we have
\beq
\delta_r = - \frac{\delta_m}{2 \delta_m + 1},
\label{deltar1}
\eeq
when we have a rapid mass 
loss with a shorter ejection timescale than the dynamical timescale.
In this rapid mass loss case, when about a half of total mass is lost 
($\delta_m \sim -0.5$), the radius expansion can be significantly 
larger than in adiabatic changes, when the mass loss occurs on a 
timescale longer than the dynamical timescale.
In this case, the expansion proceeds at a rate proportional to the 
mass loss rate ($\delta_r = - \delta_m/(\delta_m + 1)$).

In the top panel of Figure~\ref{fig:re}, we show the temporal 
evolution of half-mass radii of the simulated galaxies of No-AGN 
and two AGN feedback models: Th-AGN and MR-AGN with three 
feedback efficiencies respectively. After the final galactic nuclei 
coalesce at $t\sim1.5$ Gyr, MR-AGN-05 model has a large and 
continuous increase of the galaxy size, with the half-mass radius 
increasing by a factor of 1.25. MR-AGN models tend to have larger 
increment in size compared to the Th-AGN models. As discussed 
above, the momentum-based mechanical AGN feedback model can more 
rapidly and impulsively remove a large amount of cold gas from the 
baryon-dominated central regions of the galaxies. It can trigger a 
puffing up of the central region of the stellar component while the 
dark matter halo extending far beyond the stellar distribution 
stabilizes the system and prevents its total disruption.
However, the MR-AGN-05 model which shows the 
considerable size increase has much larger black hole
mass growth with $\Mbh > 10^9 \Msun$, which is inconsistent with 
the observed $\Mbh - \sigma$ relation. For MR-AGN-20 model 
consistent with the $\Mbh - \sigma$ relation, the size increase is negligible.
Adiabatic expansion due to mass loss from AGN feedback plays
a minor role for galaxy size growth as found in \cite{2010MNRAS.401.1099H}.

In the bottom panel of Figure~\ref{fig:re}, we show the adiabatic invariant 
of the stellar component of the system. In most models, the 
adiabatic invariant shows minor increment, while MR-AGN-05 with 
the largest size increase shows the largest increase in the adiabatic
invariant.
The puffing-up process of the mechanical feedback models is 
non-adiabatic, and more efficient than the thermal feedback 
models following Equation~\ref{deltar1}. 

\begin{figure}
\epsfig{file=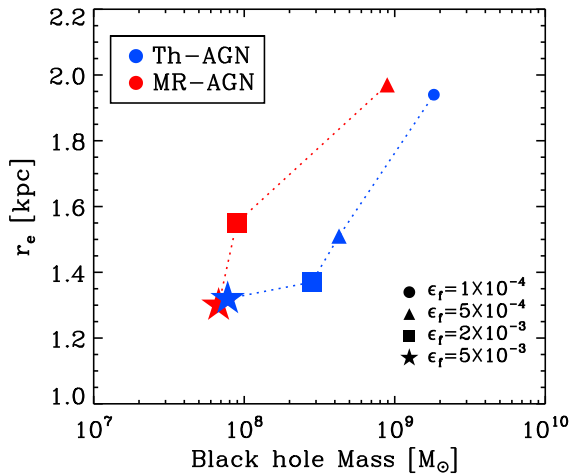,width=\columnwidth}
\caption{Galaxy size evolution of two feedback models, Th-AGN 
(blue) and MR-AGN (red), as a function of total black hole growth. 
Three models are shown for MR-AGN, with the feedback efficiency 
$\epsw = 5 \times 10^{-3}$, $2 \times 10^{-3}$ and $5 \times 10^{-4}$, 
with different symbols for the respective efficiencies. Th-AGN models with the same efficiency are shown 
along with the additional model of $\epsw = 1 \times 10^{-4}$ which 
has high enough mass growth in black hole comparable to 
MR-AGN-05.
\label{fig:growth}}
\end{figure}

We compare the effect of two AGN feedback models on the galaxy size
as a function of total black hole mass growth in Figure~\ref{fig:growth}. 
We show 3 models with MR-AGN feedback: MR-AGN-50, 
MR-AGN-20, and MR-AGN-05. In case of Th-AGN feedback 
models, we show the results with the identical feedback efficiencies, and 
additionally add one more model with the $\epsw = 1 \times 10^{-4}$ 
which shows the large enough mass growth in black hole  
comparable to MR-AGN-05. For the same mass growth in black 
hole, MR-AGN models have bigger effective radius than Th-AGN 
models as rapid mass loss effectively induces the puffing-up 
process. As before, the models with the lower AGN feedback 
efficiency show the considerable size increase but have much higher black hole
mass growth, inconsistent with the observed $\Mbh - \sigma$ relation.
Adiabatic expansion requires significant amount of mass loss by outflowing gas,
thus requires a large black hole mass growth to be the explanation 
for the observed size evolution.

\subsection{Central density}
\begin{figure}
\epsfig{file=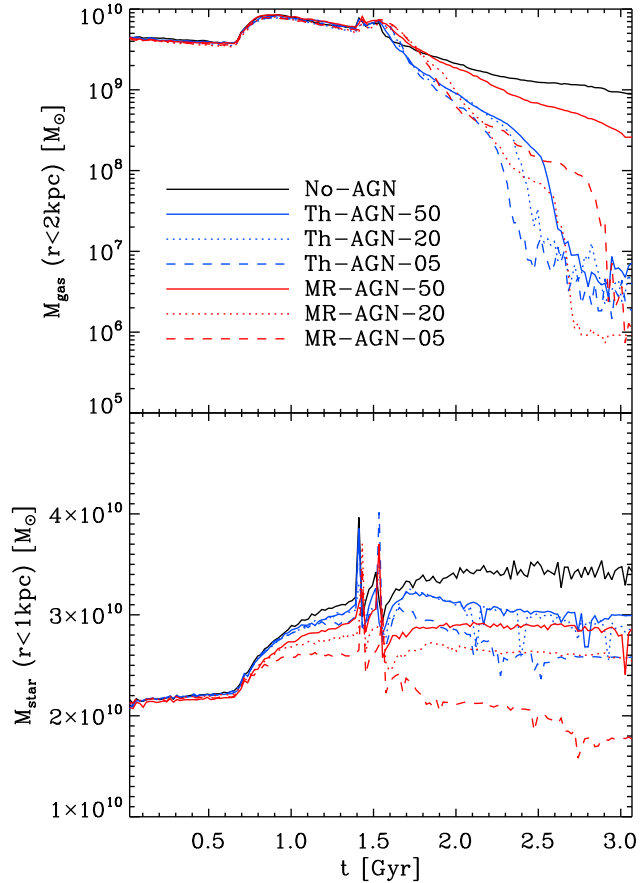,width=\columnwidth}
\caption{Evolution of gas mass enclosed within central 2 kpc radius (top) 
and stellar mass enclosed within central 1 kpc radius (bottom).
\label{fig:mass}}
\end{figure}

Figure~\ref{fig:mass} shows the evolution of the gas and the stellar 
masses contained within a fixed radius (1 kpc) in major merger 
simulations. The amount of gas in the central region of the 
galaxies decreases due to the star formation, black hole mass
accretion and the outflowing gas by the AGN feedback in both 
feedback models. The model MR-AGN-50, which has a smallest 
mass growth in black hole, shows the smallest decrease in gas 
within 2 kpc radius. In case of the stellar mass, both feedback 
models have lower central stellar density compared to No-AGN 
model mainly due to the less gas remaining in the central region
of the galaxy, but the difference tends to be larger in MR-AGN models.
Compared to the No-AGN model, Th-AGN-50 has 13 percent 
less stellar mass and MR-AGN-20 has 25 percent less stellar mass 
within 1 kpc radius. The central densities show
little decrease after the merger despite the fact that the gas is
thrown out by AGN feedback in most models. However, in 
case of MR-AGN-05 which shows the largest 
increase in galaxy size, the central density decreases by 30  percent after
the merger. As before, the quasar activity can lower the central 
stellar density of the galaxy, but only when it has enough black 
hole mass growth and a corresponding large amount of 
outflowing gas. However, the required black hole mass 
growth to explain the recent observations of the mild decrease
of central stellar mass in elliptical galaxies \citep[e.g.][]{2001ApJ...563...34M,
2010ApJ...714L.244S,2012MNRAS.422.3107S,2013ApJ...771...85V} 
is inconsistent with the observed $\Mbh - \sigma$ relation.

\section{Summary}\label{summary}
We have investigated the effects of radiative and momentum-based 
mechanical AGN feedback on the gas flow in galaxy mergers, with 
the aid of three-dimensional SPH simulations. Our numerical black 
hole model incorporates both radiative and mechanical AGN 
feedback and renders a physically more accurate picture of how a 
galaxy and its embedded black hole evolve under each others 
influence, providing a powerful tool in understanding the coevolution
of black holes and galaxies. Our main results are as follows:

\begin{itemize}
\item[1.] {\it Self-regulated black hole growth.}
We show that our AGN feedback treatment is an 
effective mechanism for halting further growth of the black hole once 
it has reached a critical size for the gravitational potential of the 
bulge. We show the successful treatment of the mechanical and 
radiation feedback in recovering the observed $\Mbh - \sigma$
relationship between the black hole mass and the galaxy velocity
dispersion with the adopted set of parameters, the initial wind velocity $\vw=$ 10,000 $\kms$ and 
 the feedback efficiency $\epsw=2 \times 10^{-3}$. This was also obtained in the previous thermal 
feedback treatments \citep[e.g.][]{2005Natur.433..604D,2005MNRAS.361..776S} and remains as a 
strong argument in helping to understand the observed physical 
relation between black hole and galactic properties. 
\item[2.]{\it Large fluctuation level in black hole mass accretion.}
In the mechanical feedback model, the fluctuation level in black
hole mass accretion, and therefore also in radiant output, is
significantly greater than the thermal feedback prescription.
Episodic accretion is the norm with bolometric luminosity
fluctuating between $\sim 10^{-1}$ and $10^{-6}$ of $L_{\rm Edd}$
during merger events.
\item[3.]{\it Galactic outflow.}
We show that our feedback model can drive large-scale galactic 
outflows, which unbind a significant fraction of the gas of the host 
galaxy. The AGN-driven winds found in this study provide a 
promising explanation for the moderate velocity outflows observed 
in some post-starburst galaxies and for the narrow-absorption line 
winds with $v \sim 500-1500$ $\kms$ seen in local quasars. 
This behaviour is consistent with the recent founding 
by \cite{2012MNRAS.420.2221D}
who also use a momentum based feedback implementation.
Outflowing kinetic energy is 20 times larger in the mechanical
feedback models than in the thermal feedback models which show
negligible effect on gas properties due to the instantaneous
cooling of the thermal energy deposited to the star forming gas in the
multiphase star formation model.  The mechanical feedback models
will have a corresponding larger effect on the
surrounding intergalactic medium \citep[e.g.,][]{2000MNRAS.318L..65F}.
\item[4.] {\it X-ray luminosity from hot gas consistent with 
observations.}
We show that the thermal feedback model with the feedback 
efficiency  of the standard value $\epsw=5 \times 10^{-3}$ appears 
to produce X-ray luminosities that are too high for their 
corresponding velocity dispersions. On the other hand, our fiducial 
model with mechanical and radiative feedback prescription spreads 
the halo gas over a larger volume and results in lower X-ray 
luminosity from the hot gaseous halo compared to the thermal 
feedback, and our proposed model with $\epsw=2.0 \times 10^{-3}$
reproduces observed $L_X - \sigma$ relation and $\Mbh - \sigma$ 
relation simultaneously.
\item[5.] {\it The effects of AGN feedback driven mass loss on the size and central density of the host galaxy.}
We show that AGN-driven mass loss can moderately increase the galaxy size
and decrease the central density. However, as noted in Section~\ref{sec:size},
the required black hole mass growth to fully account for the observed
galaxy size evolution is much larger than that observed. Given observational 
constraints on $\Mbh - \sigma$ relation, the effect of the mass loss driven by AGN 
feedback on galaxy size is moderate with the adopted set of parameters.
\end{itemize}

\section*{Acknowledgments}
The authors would like to thank Renyue Cen, Jenny Greene, Taysun
Kimm, Silvia Pellegrini, James Stone, and Michael Strauss for 
helpful conversation. We would also 
like to thank the anonymous referee for careful reading of the 
manuscript and valuable comments and suggestions that greatly 
improved this manuscript. E.C. and J.P.O. acknowledge the support of 
NSF grant AST-0707505. E.C. and T.N. acknowledge the support 
from the DFG cluster of excellence ``Origin and Structure of the 
Universe''.  E.C. was supported by the Samsung Scholarship 
foundation and made extensive use of the computing facilities of 
the Princeton Institute of Computational Science and engineering. 
P.H.J. acknowledges the support of the Research Funds of the 
University of Helsinki.

\bibliography{references}

\end{document}